\documentclass[letter]{aa} 

\usepackage{graphicx}
\usepackage{txfonts}
\usepackage[colorlinks]{hyperref}
\newcommand{\gr}{$\gamma$-ray}
\newcommand{\nup}{$\nu_{\rm peak}^S$}
\newcommand{\fermi}{{\it Fermi}}

\newcommand{\swift}{{\it Swift}}

\newcommand{\ergs}{erg cm$^{-2}$ s$^{-1}$}
\newcommand{\erg}{erg s$^{-1}$}
\newcommand{\vareps}{\varepsilon}
\usepackage{lineno}
\usepackage{ulem}
\newcommand{\swiftdeepsky}{\textit{Swift-DeepSky}} 
\newcommand{\swiftxrtproc}{\textit{Swift-xrtproc}} 
\newcommand{\HSPJ}{3HSP\,J095507.9+355101\xspace}
\newcommand{\TXS}{TXS\,0506+056\xspace} 

\newcommand{\lsim}{{\lower.5ex\hbox{$\; \buildrel < \over \sim \;$}}}
\newcommand{\gsim}{{\lower.5ex\hbox{$\; \buildrel > \over \sim \;$}}}

\begin{document}
\title{3HSP J095507.9+355101: a flaring extreme blazar coincident in space and time 
with IceCube-200107A}
   \author{
          P. Giommi
          \inst{1,2,3}
          \and
          P. Padovani
          \inst{4,5}
          \and 
          F. Oikonomou
          \inst{4,6,7}
          \and
          T. Glauch\inst{1,6}
          \and
          S. Paiano\inst{8,9}
          \and 
          E. Resconi\inst{6}
          }

   \institute{
Institute for Advanced Study, Technische Universit{\"a}t M{\"u}nchen, 
Lichtenbergstrasse 2a, D-85748 Garching bei M\"unchen, Germany
               \and Associated to Agenzia Spaziale Italiana, ASI, via del Politecnico s.n.c., I-00133 Roma, Italy 
                \and ICRANet, Piazzale della Repubblica 10, I-65122, Pescara, Italy
                 \and European Southern Observatory, Karl-Schwarzschild-Str. 
2, D-85748 Garching bei M\"unchen, Germany
            \and Associated to INAF - Osservatorio Astronomico di Roma, via Frascati 33,
I-00040 Monteporzio Catone, Italy
           \and Technische Universit{\"a}t M{\"u}nchen, Physik-Department, 
James-Frank-Str. 1, D-85748 Garching bei M{\"u}nchen, Germany
            \and Institutt for fysikk, NTNU, Trondheim, Norway
           \and INAF - Osservatorio Astronomico di Roma, via Frascati 33, I-00040, Monteporzio Catone, Italy 
           \and INAF - IASF Milano, via Corti 12, I-20133, Milano, Italy \\
                  \email{giommipaolo@gmail.com}
}





\abstract
{The uncertainty region of the  highly energetic neutrino IceCube200107A includes 3HSP J095507.9+355101 ($z$~=~0.557), an extreme blazar,
which was detected in a high, very hard and variable X-ray state shortly after the neutrino arrival.
Following a detailed multi-wavelength investigation, we confirm that the source is a genuine BL Lac, contrary to TXS\,0506+056, the first source so far associated with IceCube neutrinos, which is a ``masquerading'' BL Lac. As in the case of TXS\,0506+056, 3HSP J095507.9+355101 is also way off the so-called ``blazar sequence''. We consider 3HSP J095507.9+355101 a possible counterpart to the IceCube neutrino. Finally, we discuss some theoretical implications in terms of neutrino production.
}


\titlerunning{3HSP J095507.9+355101}
\authorrunning{Giommi, P. et al.}
\maketitle
 
\section{Introduction}\label{sec:Introduction}

\begin{figure}
\hspace*{-1.cm}
\vspace*{-.3cm}
\includegraphics[width=0.58\textwidth]{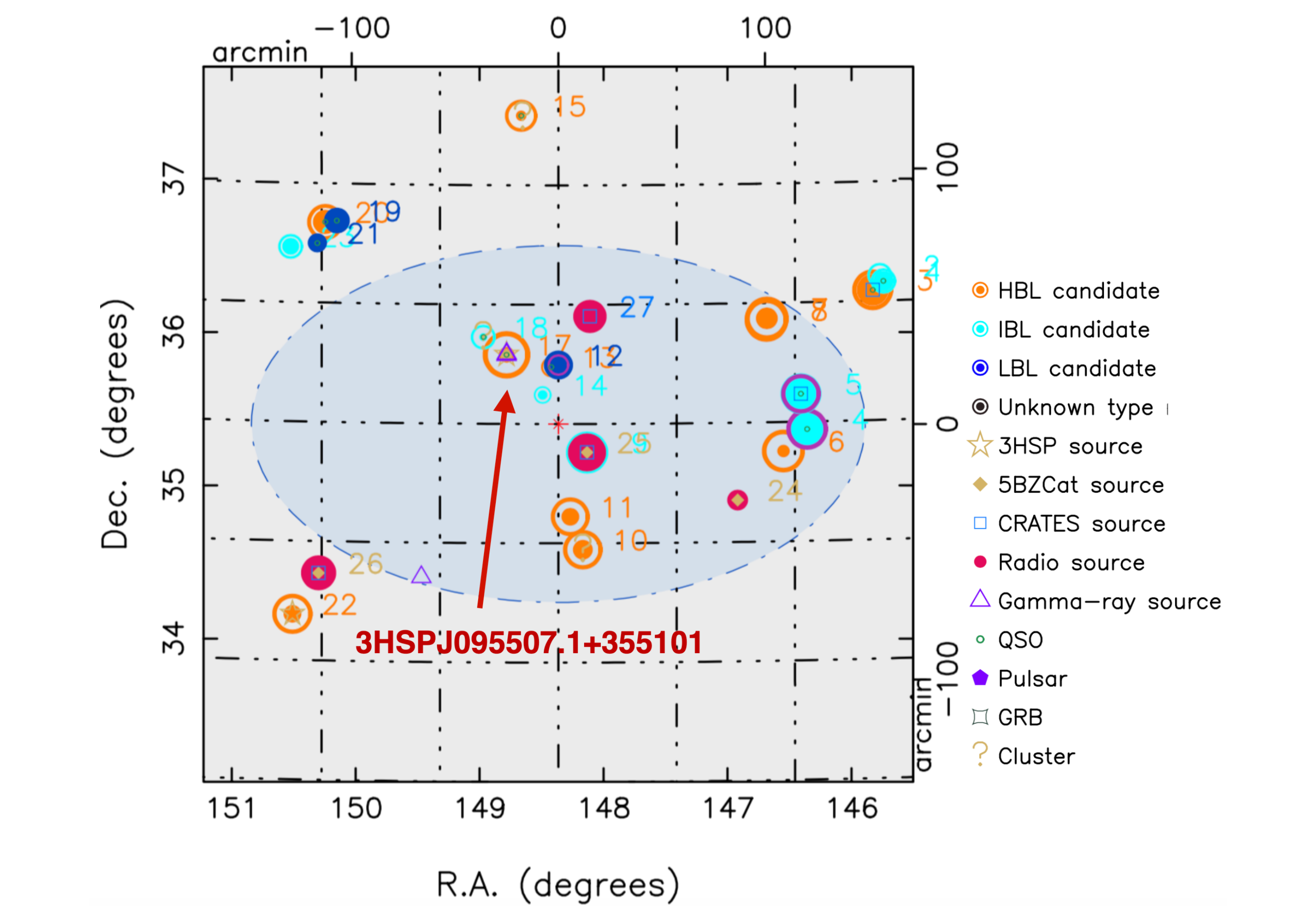}
\caption{Known and candidate blazars (radio/X-ray matching sources) around the 90\% containment region of IceCube200107A, approximated by the darker elliptical area. 
}
\label{fig:region}
\end{figure}

The IceCube Neutrino Observatory at the South Pole\footnote{\url{http://icecube.wisc.edu}} 
has detected tens of high-energy neutrinos of likely astrophysical origin \citep[e.g.][and references therein]{ICECube17_2,2019ICRC...36.1004S,2019ICRC...36.1017S}. 
So far, only one astronomical object has been significantly associated (in space and time) with some of 
these neutrinos, i.e., the bright blazar TXS\,0506+056 at $z=0.3365$ \citep{icfermi,iconly,dissecting,paiano2018txs}. It is clear, however, that blazars cannot be responsible for the whole IceCube signal (see   \citealt{ICECube17_3} and \citealt{2017ApJ...835...45A}). 
The case for some blazars being neutrino sources, however, is mounting. 
Several studies have reported hints of a correlation between blazars and 
the arrival direction of astrophysical neutrinos 
(e.g. \citealt{Pad_2014,Padovani_2016,Lucarelli_2019} and references therein) and possibly
of Ultra High Energy Cosmic Rays \citep{Resconi_2017}. Moreover, 
very recently some of the authors of this paper \citep{Giommi_2020} have extended the detailed dissection of the region around the IceCube-170922A event
related to TXS\,0506+056 carried out by \cite{dissecting}
to all the 70 public IceCube high-energy neutrinos that are well reconstructed (so-called tracks) and off the Galactic plane.
This resulted in a $3.23\,\sigma$ (post-trial) excess of IBLs\footnote{Blazars are divided based on the
rest-frame frequency of the low-energy (synchrotron) hump (\nup) into LBL/LSP
sources (\nup~$<10^{14}$~Hz [$<$ 0.41 eV]), intermediate-
($10^{14}$~Hz$<$ \nup~$< 10^{15}$~Hz [0.41 eV -- 4.1 eV)], and
high-energy (\nup~$> 10^{15}$~Hz [$>$ 4.1 eV]) peaked (IBL/ISP and HBL/HSP) sources respectively \citep{padgio95,Abdo_2010}.} and HBLs 
with a best-fit of $15 \pm 4$ signal sources, while no excess was found for LBLs. 
Given that TXS\,0506+056 is also a blazar of the IBL/HBL type \citep{Padovani_2019} 
this result, together with previous findings, consistently points to growing evidence 
for a connection between some IceCube neutrinos and IBL and HBL blazars.
We report here on 3HSP J095507.9+355101, an HBL within the error region of the IceCube
track IceCube-200107A (see Fig. \ref{fig:region}), which was found to exhibit an X-ray flare the day after the neutrino 
arrival. This source belongs to the third high-synchrotron peaked (3HSP) catalogue \citep{3hsp},
which includes blazars with \nup~$> 10^{15}$~Hz. Actually, with a catalogued synchrotron peak frequency of $\sim 5\times 10^{17}$ Hz, and a significantly higher value during the flare (Sec. \ref{subsec:Swift}),
this source belongs to the rare class of extreme blazars \citep[e.g.][and references therein]{Biteau_2020}. 
We also comment on the nature of the source and on the theoretical implications in terms of neutrino production. 
We use a $\Lambda$CDM cosmology with $H_0 = 70$ km s$^{-1}$ Mpc$^{-1}$, $\Omega_{\rm m,0} = 0.3$, 
and $\Omega_{\Lambda,0} = 0.7$.

\section{Multi-messenger data}
\subsection{IceCube data}
\label{subsec:IC}
On January 7, 2020 the IceCube Collaboration reported the detection of a high-energy neutrino candidate \citep[HESE,][]{Stein_2020} of possible astrophysical origin. While the event was not selected by the standard real-time detection procedure, it was identified as a starting track by a newly developed deep neural-network event classifier \citep{dnn}. After applying off-line reconstructions the arrival direction is given as right ascension $148.18^{+2.20}_{-1.83}$ deg and declination $35.46^{+1.10}_{-1.22}$ deg at 90\% C.L. As this was an unscheduled report, IceCube does not provide any energy information. Assuming an $E^{-2}$ spectrum and the effective area for HESE starting tracks~\citep{2014PhRvL.113j1101A}, at this declination 90\% of neutrinos have energy $0.33^{+2.23}_{-0.27}$ PeV\footnote{For an assumed $E^{-1}/E^{-2.7}$ neutrino spectrum, 90\% of neutrinos have energy $1.40^{+5.75}_{-1.22}$~PeV /$0.16^{+0.83}_{-0.12}$~PeV respectively.}. Further hints for the event being astrophysical comes from the direction. Being clearly up-going, an atmospheric muon origin can be excluded. Also the fraction of the conventional atmospheric muon neutrino background is suppressed compared to the horizon. In a follow-up GCN report \citep{Pizzuto_2020} IceCube announced the detection of two additional neutrino candidates in spatial coincidence
with the 90\% containment region of IceCube-200107A in a time range of two days around the alert and consistent with atmospheric background at a 4\% level.
Note that the error region of IceCube-200107A is also fully inside the $16.5^{\circ}$ median angular error circle of a HESE shower detected by IceCube in 2011 (HES9), and reported in \cite{2014PhRvL.113j1101A} 
In fact 3HSP J095507.9+355101 is located only 0.62$^{\circ}$ and 2.73$^{\circ}$ away from the best-fit position of IceCube-200107A and HES9, respectively.

 We estimate the flux required to detect, on average, one muon neutrino with IceCube at a specified time interval, $\Delta T$ by assuming the neutrino event, IceCube-200107A, to be a signal event. The number of signal-only, muon (and antimuon) neutrinos detected during $\Delta \mathrm{T}$ at declination $\delta$ is given by $N_{\nu_{\mu}}~=~\int_{E_{\nu_{\mu},{\rm min}}}^{E_{\nu_{\mu},{\rm max}}} {\rm d} E_{\nu_{\mu}} A_{\rm eff}(E_{\nu_{\mu}},\delta) \phi_{E_{\nu_{\mu}}} \Delta \mathrm{T},$
where $E_{\nu,{\rm min}}$ and $E_{\nu,{\rm max}}$, are the 90\% C.L. lower and upper limits on the energy of the neutrino respectively, $A_{\rm eff}$ is the effective area, and $\phi_{E_{\nu_{\mu}}}$ the muon neutrino differential energy flux. We assume a source emitting an $E^{-2}$ neutrino spectrum between 65 TeV and 2.6 PeV, the energy range in which we expect 90\% of neutrinos detected from the direction of IC-200107A in the HESE channel. Since the neutrino emission duration is unknown we calculate the neutrino flux needed to produce one neutrino in IceCube from the direction of IceCube-200107A for $\Delta \mathrm{T}$ = 30 d/250 d/10 yr, corresponding to the lower limit on the duration of the UV/soft X-ray flare (Sec.~\ref{subsec:Swift}), the 
\fermi~integration time (Sec.~\ref{subsec:Fermi}) and the duration of the IceCube operation respectively. Using the effective area of \cite{Blaufuss}, we obtain an integrated all-flavour neutrino energy flux, of $ 3 \times 10^{-9} / 4 \times 10^{-10}/ 3 \times 10^{-11}$~\ergs ~respectively, for a source at $\delta = 35.46^{\circ}$. This corresponds to  energy-integrated, all-flavour neutrino luminosity, in the central 90\% energy range, of $\mathcal{L}_{\nu}\approx 4 \times 10^{48} / 5 \times 10^{47}/ 3 \times 10^{46}~{\rm erg~s^{-1}}$  for a source at $z = 0.557$. For a population of neutrino producing sources with summed expectation of order one neutrino, the energy flux estimate given above roughly corresponds to the total energy flux produced by the source population, whereas the individual source contribution, and thus the individual neutrino luminosity is much lower than our estimate above~\citep{icfermi}. 

\begin{figure}
\includegraphics[width=0.49\textwidth]{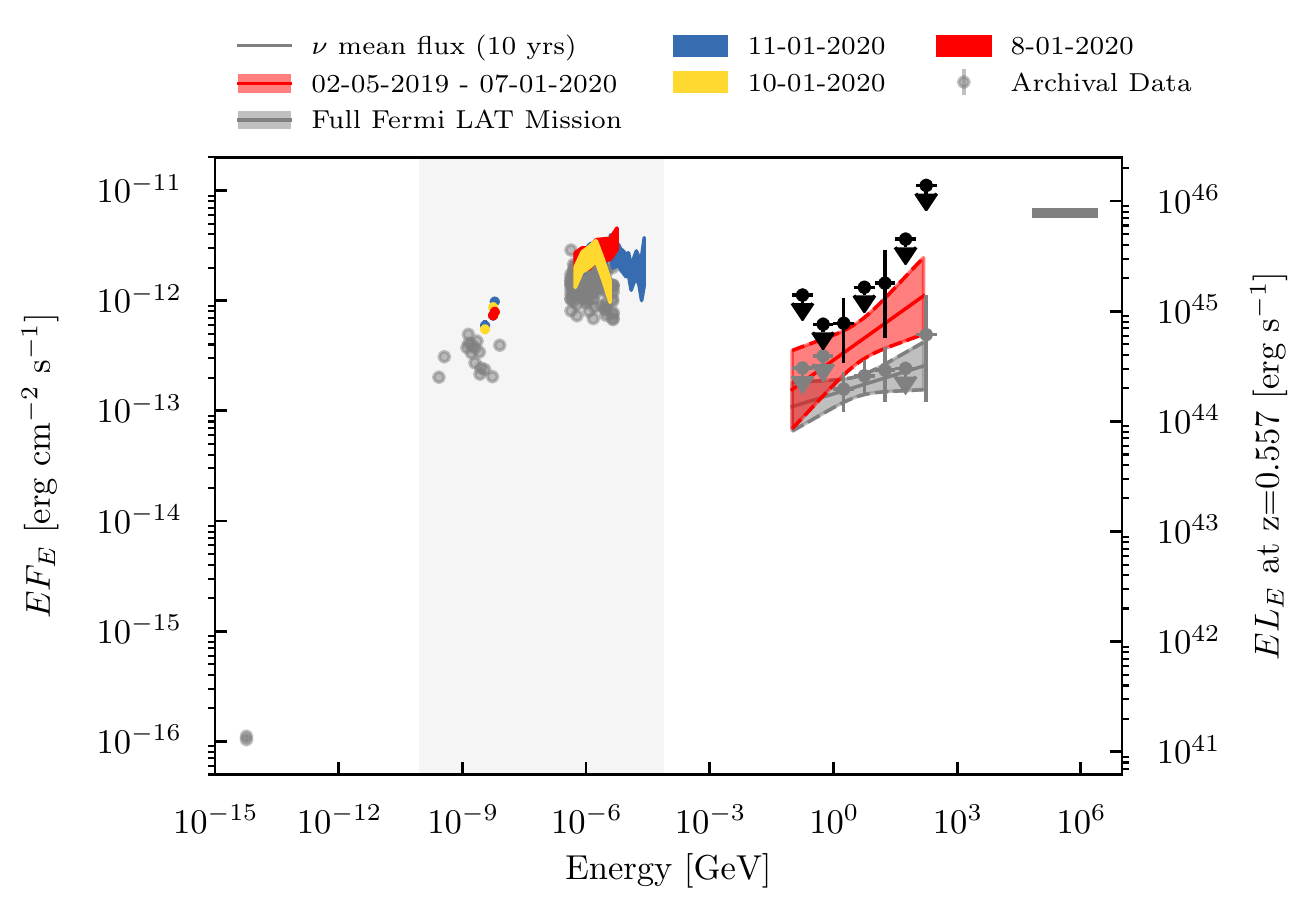}
\includegraphics[width=0.48\textwidth]{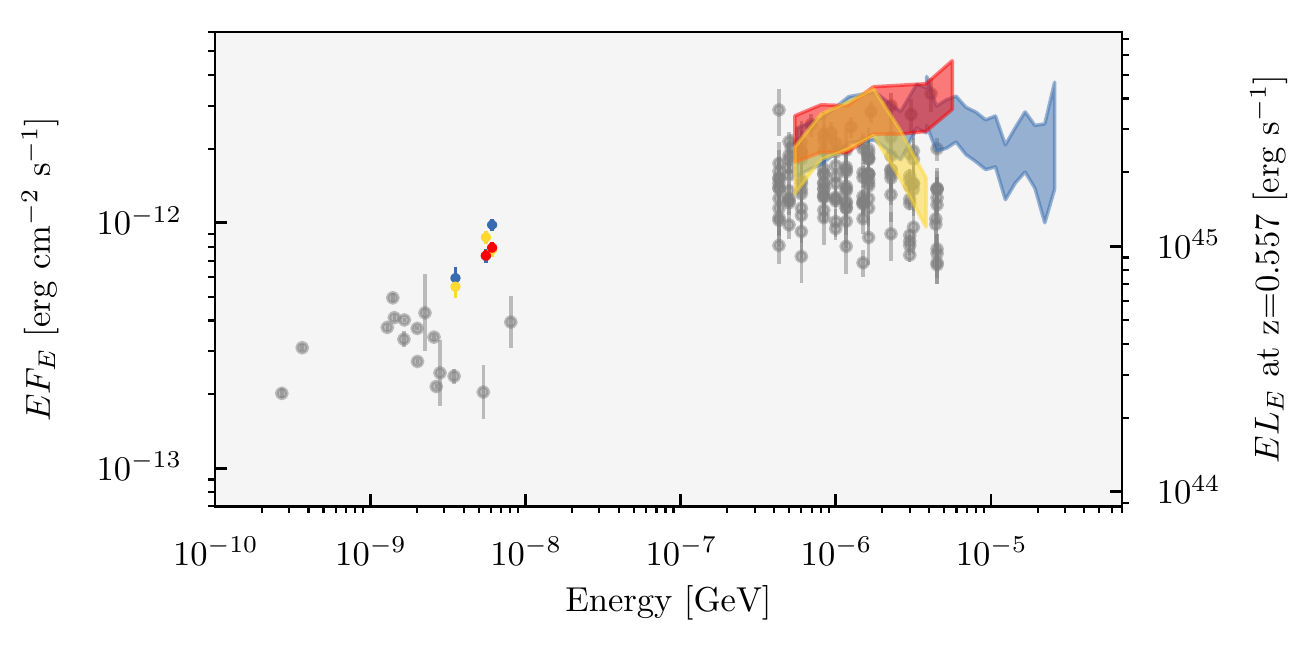}
\vspace{-0.2cm}
\caption{The spectral energy distribution (SED) of 3HSP J095507.1+355101: grey points refer to archival data and, in the case of {\it Fermi}-LAT data, the time-integrated measurement up to the arrival of the neutrino alert. The mean all-flavour neutrino flux is shown for an assumed live time of 10 yr. 
Coloured data-points are Swift and NuSTAR measurements made around the neutrino arrival time.
 The black/grey \gr\ points refer to the red/grey bow-ties, indicating the one-sigma uncertainty of the \gr\ measurement 250 days before the observation of IceCube-200107A and during the full mission, respectively. The best-fit fluxes are shown as solid lines. The upper panel shows the full hybrid SED while the bottom one provides an enlarged view of the optical and X-ray bands.}
\label{fig:SED}
\end{figure}
\begin{figure}
\includegraphics[width=0.5\textwidth]{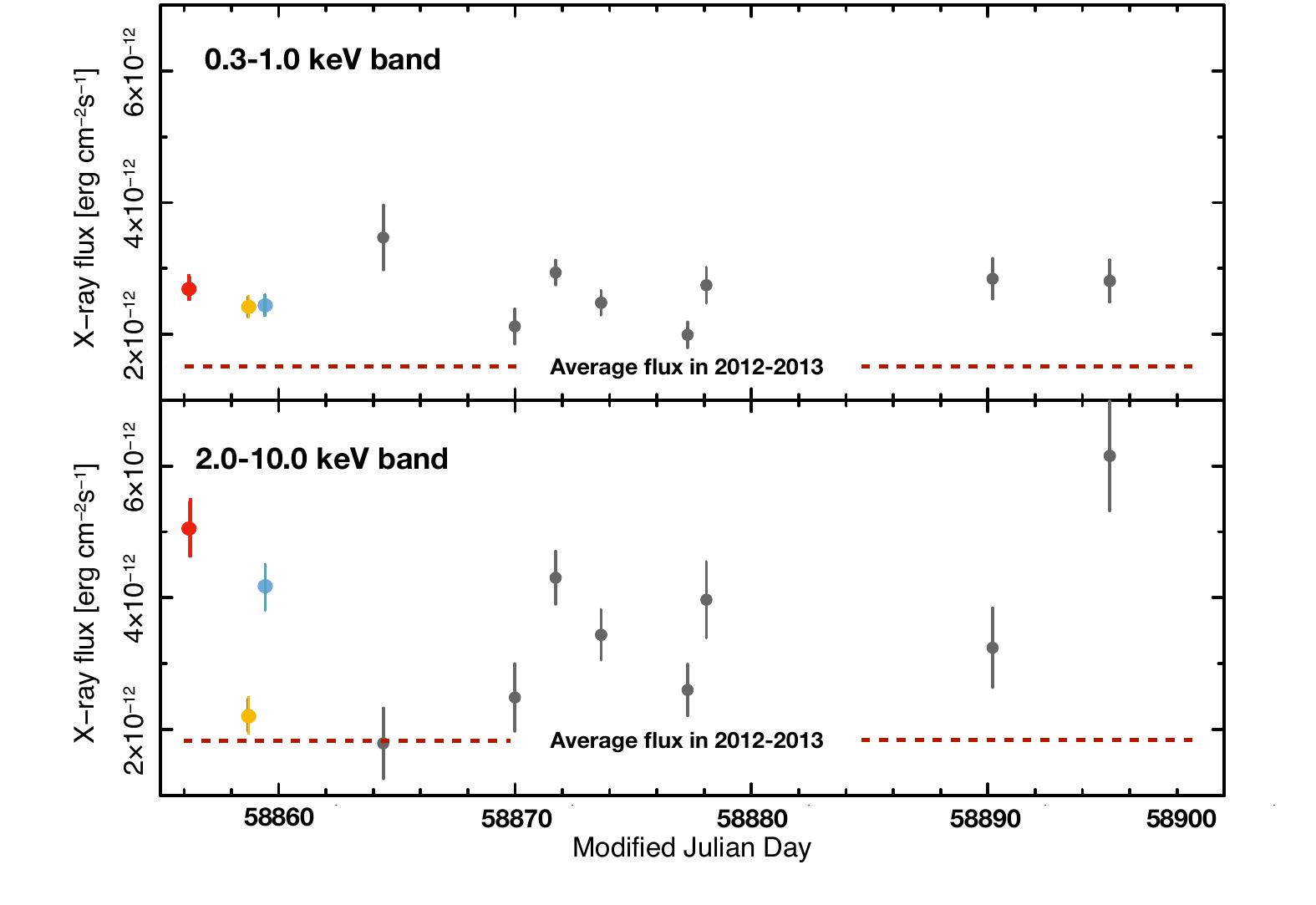}
\vspace{+0.2cm}
\vspace{-0.5cm}
\caption{Swift soft and hard X-ray monitoring of 3HSPJ095507.1+355101 after the neutrino arrival. The first observation was carried out one day after the detection of IC200107A. Colours match the ones used in the SED of Fig. \ref{fig:SED}. Note that the flux is higher than the average  observed in 2012-2013 in both bands, but short-term variability is only present in the $2-10$ KeV energy band.}
\label{fig:LCX}
\end{figure}

\subsection{{\it Swift} data}
\label{subsec:Swift}
The {\it Neil Gehrels Swift} observatory \citep{swift} observed 3HSP J095507.9+355101 37 times; 26 pointings were performed between 2012 and 2013, and the remaining ones have been carried out either as a Target of Opportunity (ToO), after the IceCube200107A event, which revealed the source to be in a flaring and very hard state \citep{GiommiAtel,KraussAtel} and as part of a subsequent monitoring program.
We analysed all the X-Ray Telescope \citep[XRT;][]{xrt} imaging data using \swiftdeepsky, a Docker container\footnote{\url{https://www.docker.com}}
encapsulated pipeline software developed in the context of the Open Universe initiative \citep{openuniverse,Giommi2019}.
Spectral analysis was also performed on all exposures with sufficiently strong signal using the XSPEC-12 software embedded in a dedicated processing pipeline, called \swiftxrtproc, first presented in \cite{Giommi2015}.
Details of the results are given in the Appendix.
The $2-10$ keV emission from 3HSP J095507.9+355101 exhibited over a factor of ten variability in intensity associated with strong spectral changes following a harder-when-brighter trend (see Tab. \ref{imaginggresults},  \ref{spectralresults}, and Fig.\ref{fig:flvsslope}).
The ToO observation of 3HSP J095507.9+355101 found this object in a flaring and hard state, with a 2 -- 10 keV X-ray flux of $\sim 5\times10^{-12}$ \ergs\, a factor 2.5 larger than the average value observed in 2012 -- 2013, and with a power law spectral index  $\Gamma =1.8 \pm 0.06$. A log-parabola model 
gives a similar slope at 1 keV and curvature parameter consistent with zero, implying \nup $\gsim 2 \times 10^{18}$ Hz.
The optical and UV data of the Ultra-Violet and Optical telescope \citep[UVOT;][]{Roming} were analyzed using the on-line tools of the SSDC interactive archive\footnote{\url{http://www.ssdc.asi.it}}. 
Spectral data are shown in Fig. \ref{fig:SED}, while the X-ray light-curve is shown in Fig. \ref{fig:LCX}.
The optical/UV and low energy X-ray data reach their maximum intensity after the neutrino arrival and remain approximately constant for the subsequent $\sim 30$ days, implying that all the variability in the $2-10$ keV band is induced by strong spectral changes above $\sim 7.3 \times 10^{17}$ Hz.

\subsection{NuSTAR data}

3HSP J095507.9+355101 was observed by the {\it NuSTAR} hard X-ray observatory \citep{nustar} four days after the detection of IceCube-200107A, following the results of the \swift\ ToO mentioned above. The observation was partly simultaneous with the third \swift\ pointing after the neutrino event.  The source was detected between 3 and $\sim 30$ keV. A power law spectral fit gives a best fit slope of $\Gamma = 2.21 \pm 0.06$ with a reduced $\chi^2_{\nu} = 0.93$. 
The data, converted to SED units, are shown as light blue symbols in Fig. \ref{fig:SED}.

\subsection{{\it Fermi}-LAT data}
\label{subsec:Fermi} 
The analysis of the $\gamma$-ray emission of 3HSP J095507.9+355101 is based on publicly available Fermi-LAT Pass 8 data acquired in the period August 4, 2008 to January 8, 2020.  In order to describe the spectral evolution of the source two time windows are analysed, the full mission and the last 250 days before the detection of IceCube-200107A. The 250 days are needed in order to ensure the collection of sufficient photon statistics. 
The resulting fit between MJD 58605.6 and 58855.6 gives an evidence for emission with significance of $2.9 \sigma$ and spectral index of $\Gamma = 1.73\pm 0.31$ for a typical single power-law model. The spectral index over the full-mission is $\Gamma=1.88 \pm 0.15$, with photon associations up to 178 GeV at $99$ \% C.L  and a detection significance of 6.3 $\sigma$. The corresponding photon fluxes integrated over the entire energy range between 100 MeV and the highest-energy photon at 178 GeV are $(1.11^{+0.95}_{-0.52})\times 10^{-9}\,{\rm ph\,cm^{-2}\,s^{-1}}$ (250 days) and $(0.61^{+0.27}_{-0.19})\times 10^{-9} \,{\rm ph\, cm^{-2}\,s^{-1}}$ (full-mission), respectively. The best-fit spectra are also visualized together with their respective SED points in Figure \ref{fig:SED}. More details on the data analysis are given in the Appendix.

\subsection{LBT data}

3HSP~J095507.9+355101 was observed on January 29, 2020 at the Large Binocular 
Telescope (LBT; \citealt{pogge2010}) in the optical band ($4,100 - 8,500~\textrm{\AA}$). A firm redshift z~=~0.557 was derived 
thanks to the clear detection of absorption features attributed to its host galaxy. No narrow emission line was detected down 
to an equivalent width $\sim$ 0.3 $\textrm{\AA}$. This corresponds to [\ion{O}{II}] and  [\ion{O}{III}] line luminosities  
$< 2\times 10^{40}$~erg s$^{-1}$. Details about the spectroscopic study of the source, its host galaxy, and close environment are given in \cite{paiano2020}.


\section{The nature of 3HSP J095507.9+355101}

The SED of 3HSP J095507.9+355101, assembled using  multi-frequency historical data, shows that this source exhibits a \nup $\sim 5\times 10^{17}$ Hz \citep{3hsp}, a very large value that is rarely reached even by extreme blazars \citep{Biteau_2020}. The 3HSP catalogue includes only 80 sources with \nup $\geq 5\times 10^{17}$ Hz that have been detected by Fermi-LAT in $\sim $34,000 square degrees of high Galactic latitude sky (|b| > 10$^{\circ}$), corresponding to an average density of one object every 425 square degrees. The chance probability that one such extreme source is included in the 7.3 square degrees error region of IC200107A is therefore 7.3/425, or about 1.7\%. 
At the time of the neutrino detection 3HSP J095507.9+355101 was also found to be in a very hard state (\nup~$\gtrsim 2\times 10^{18}$ Hz [10 keV] and flaring; see Fig. \ref{fig:SED}). Blazars are known to spend a small fraction of their time in a very high X-ray state \citep{Giommi1990}. We used the Open Universe blazar database (Giommi et al. 2020, in preparation) to estimate how frequently the extreme sources with \nup~$\gtrsim 5\times 10^{17}$ Hz, and observed by Swift \citep[in WT or PC mode,][]{xrt} more than 100 times (MRK421, MRK501, 1ES2344+514 and 1ES0033+595), are detected in a flaring state, finding that they spend less than 10\% of the time at an intensity that is larger than twice the average value. 
The overall chance probability of finding a blazar with \nup\ as high as that of 3HSP J095507.9+355101 in the error region of IceCube-200107A during a flaring event is therefore a fraction of 1\%. Since this is a posterior estimation based on archival data, which may hide possible biases, it should not be taken as evidence for a firm association, but rather as the identification of an uncommon and physically interesting event that corroborates a persistent trend 
\citep[e.g.][]{icfermi,Giommi_2020} and motivates this work.
We have studied the nature of 3HSP J095507.1+355101, following \cite{Padovani_2019}, to check if this
source is also a ``masquerading'' BL Lac like TXS\,0506+056, i.e., intrinsically a flat-spectrum radio quasar (FSRQ) 
with the emission lines heavily diluted by a strong, Doppler-boosted jet. Given the upper limits on its 
$L_{\rm \ion{O}{II}}$ and $L_{\rm \ion{O}{III}}$ and its black hole mass estimate \citep[${\rm M}_{\rm BH} \sim 3 \times 10^{8}~ 
{\rm M}_{\odot}$;][]{paiano2020}, we obtain the following results: 1.
its radio and \ion{O}{II} luminosities put it at the very edge of the locus of jetted quasars \citep[Fig. 4
of][]{Kalfountzou_2012}; 2. its Eddington ratio is $L/L_{\rm Edd} < 0.02$, formally still within the range of high-excitation
galaxies (HEGs, characterized by $L/L_{\rm Edd} \gtrsim 0.01$) but barely so; 3. its broad-line region (BLR) power 
in Eddington units is $L_{\rm BLR}/L_{\rm Edd} < 3 \times 10^{-4}$, which implies that this source is not an FSRQ
according to \cite{Ghisellini_2011} (as this would require $L_{\rm BLR}/L_{\rm Edd} \gtrsim 5 \times 10^{-4}$); 
4. finally, its $L_{\gamma}/L_{\rm Edd}$ values range between
$\sim 0.04$ and $\sim 0.10$, depending on its state, i.e. they straddle the BL Lac -- FSRQ division proposed by 
\cite{Sbarrato_2012} ($L_{\gamma}/L_{\rm Edd} \sim 0.1$). Based on all of the above we consider 3HSP J095507.1+355101
an unlikely ``masquerading'' BL Lac.  
\begin{figure}
\includegraphics[width=0.5\textwidth]{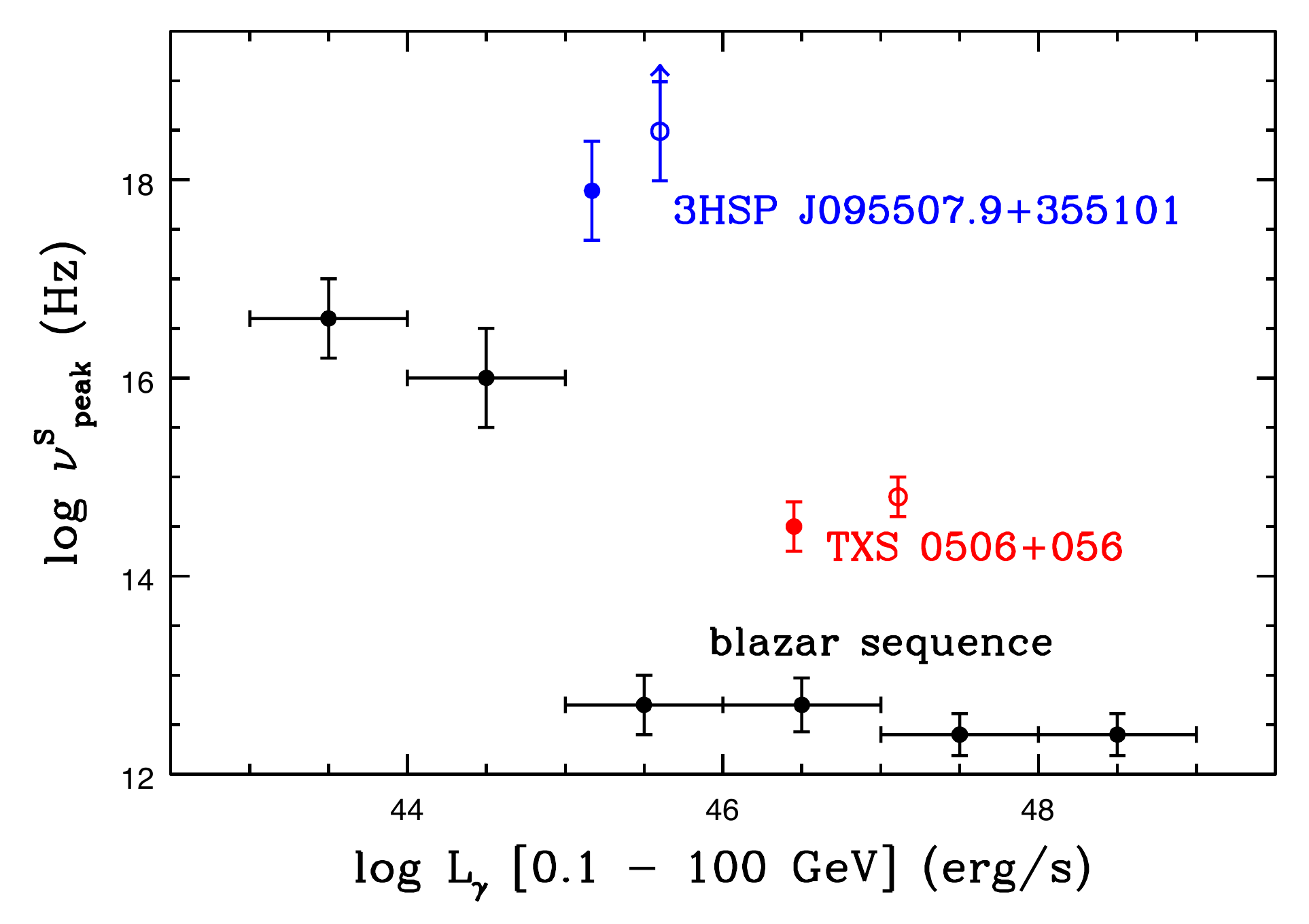}
\caption{Rest-frame \nup~versus $L_{\gamma}$ for the revised blazar sequence (black
  points; \citealt{Ghisellini_2017}) and TXS\,0506+056 and 3HSP J095507.9+355101 
  (red and blue points respectively: average
  [filled] and $\gamma$-ray flare [open] values). The TXS\,0506+056 values are from
  \protect\cite{Padovani_2019}. The error bars denote the sample dispersion (blazar
  sequence) and the uncertainty (TXS\,0506+056 and 3HSP J095507.9+355101) respectively.}
\label{fig:sequence}
\end{figure}
Fig. \ref{fig:sequence} shows the location of 3HSP J095507.1+355101 on the \nup~versus $L_{\gamma}$ plane
in its average state (blue filled point) and during the flare  (blue open point). The source is an extreme outlier of the so-called blazar sequence, even more so than TXS\,0506+056.
Given its $L_{\gamma}$, in fact, its \nup~should be 
about five orders of magnitude smaller to fit the sequence.

\section{Theoretical considerations and conclusion}

We now present some general, model-independent, theoretical constraints on neutrino production by \HSPJ based on the multi-wavelength observations. A comprehensive overview of models of neutrino emission from \HSPJ is presented in~\citep{Petropoulou2020}. Neutrino production in the blazar jet is most likely facilitated by photopion ($p\pi$) interactions. The neutrino production efficiency can thus be parametrised by $f_{p\pi}$, the optical depth to ${p\pi}$ interactions. Of the energy lost by protons with energy $\vareps_p$ in $p\pi$ interactions, $3/8$ths go to neutrinos, resulting in the production of neutrinos with all-flavour luminosity, $\vareps_{\nu} L_{\vareps_{\nu}} = (3/8) f_{p\pi} \vareps_p L_{\vareps_p}.$ Each neutrino is produced with energy $\vareps_{\nu} \approx 0.05 \vareps_{p}$. Here and throughout, $\vareps L_{\vareps}$ is the luminosity per logarithmic energy, $\vareps \cdot \mathrm{d}L/\mathrm{d}\vareps$, unprimed symbols denote quantities in the cosmic rest frame, quantities with the subscript ``{\it obs}'' refer to the observer frame, and primed quantities refer to the frame comoving with the jet. Neutrinos produced in interactions with photons comoving with the jet have typical energy~ $\vareps_{\nu,\rm obs}~\approx~7.5~{\rm PeV}\left(\vareps_t \right/2~\rm keV)^{-1}\left(\Gamma/20\right)^2(1+z)^{-2}$, where $\Gamma$ is the bulk Lorentz factor of the jet, and $\vareps_t$ the energy of the target photons assuming that protons are accelerated to at least 150~PeV. 

\begin{figure}
\includegraphics[width=0.49\textwidth]{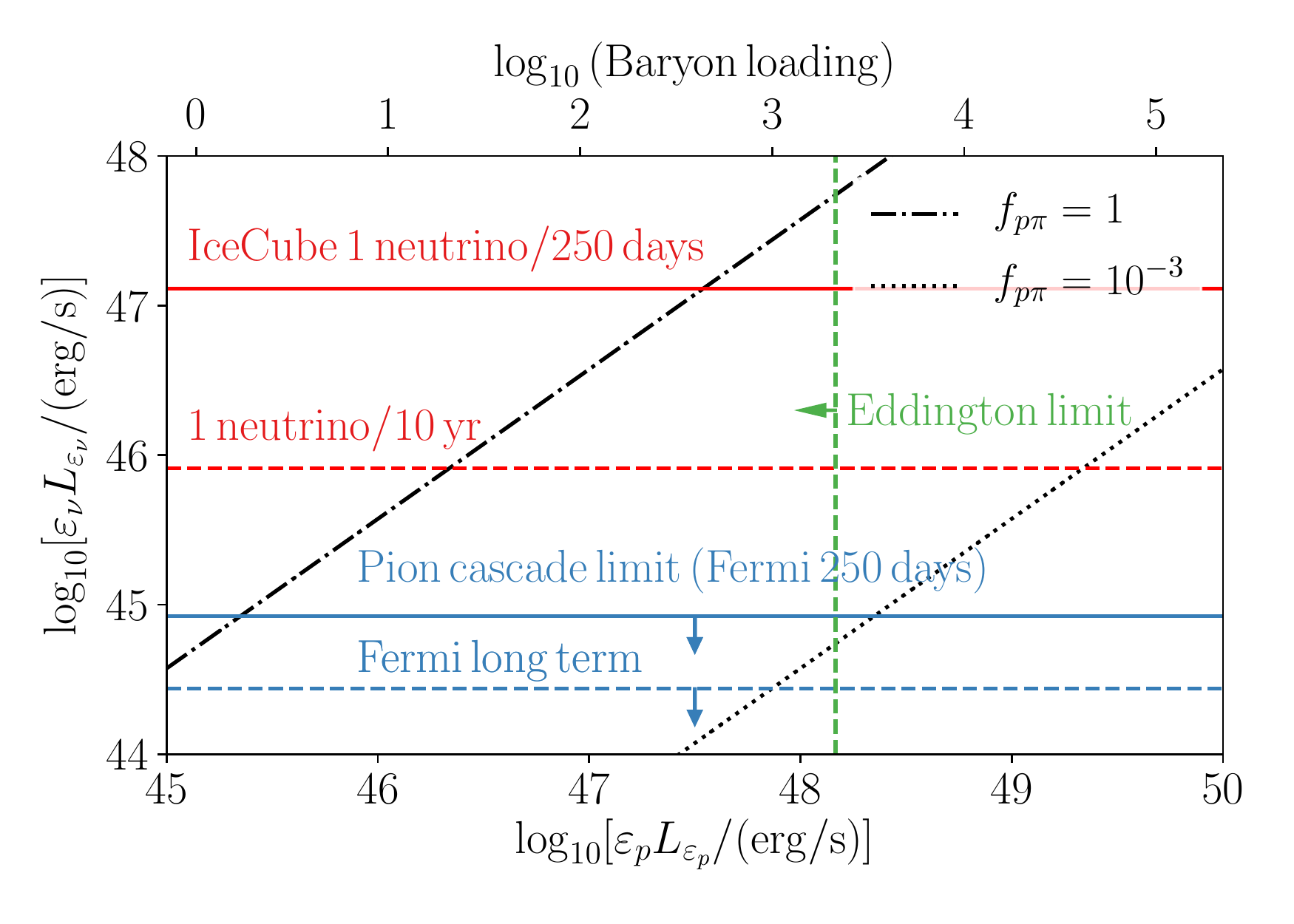}
\vspace{-0.5cm}
\caption{All-flavour neutrino luminosity as a function of proton luminosity for two different values of the optical depth to photopion interactions $f_{p\pi}$. The red solid (dashed) line gives the neutrino luminosity corresponding to 1 muon neutrino in IceCube from \HSPJ if the neutrino emission lasted 250 days (10 yr). The blue horizontal solid (dashed) line gives the upper limit to the neutrino luminosity implied by the \fermi-LAT 250-day (long-term average) spectrum. The green line shows the upper limit to the proton luminosity implied by the Eddington luminosity of the $3\times10^{8}M_{\odot}$ black hole, assuming $\Gamma = 20$, proton spectral index -2, and maximum proton energy $10^{18}$ eV.}
\label{fig:LC}
\end{figure}

The remaining 5/8ths of the proton energy lost go towards the production of electrons and pionic $\gamma$-rays. Synchrotron emission from electrons/positrons produced in $p\pi$ interactions and two-photon annihilation of the pionic $\gamma$-rays result in synchrotron cascade flux~\citep{2018ApJ...865..124M}
\begin{equation}
\tiny
\vareps_{\nu} L_{\vareps_{\nu}} \approx \frac{6 (1+Y_{\rm IC})}{5}\vareps_{\gamma} L_{\vareps_{\gamma}}|_{\vareps^{p\pi}_{\rm syn}} \approx 8 \times 10^{44}~{\rm erg~s^{-1}} \left( \frac{\vareps_{\gamma} L_{\vareps_{\gamma}}|_{\vareps^{p\pi}_{\rm syn}}}{7\times 10^{44}
}\right)
\label{eq:2}
\end{equation}
\noindent where $Y_{\rm IC}$ is the Compton-Y parameter, typically expected to be $Y_{\rm IC} \ll 1$ and the $\gamma$-ray emission is expected at energy $\vareps^{p\pi}_{\rm syn,obs} \approx 39.4~ {\rm GeV} (B/0.3~{\rm G}) (\vareps_{\nu,\rm obs} / 7.5 ~\rm PeV)^2 (20/\delta) (1+z)^{-1}$. The 250-day average luminosity of the flaring SED of \HSPJ in the \fermi-LAT energy range thus imposes a limit to the average neutrino luminosity according to Eq.~\ref{eq:2}. If the neutrino emission lasted 250 days, the expected neutrino luminosity of Eq.~\ref{eq:2}, is $\sim$ 2.2 orders of magnitude lower than the flux implied by the detection of one neutrino according to the estimate of Sec.~\ref{subsec:IC}, which is  $\vareps_{\nu} L_{\vareps_{\nu}} = \mathcal{L}_{\nu}/\ln{(2.6~\mathrm{PeV}/65~\mathrm{TeV})} \approx 1.3 \times 10^{47}$\erg.  
The expected neutrino luminosity as a function of the proton luminosity is shown in Fig.~\ref{fig:LC}, for two characteristic values of $f_{p\pi}$ (by definition $f_{p\pi} \leq 1$), together with the constraint imposed by Eq.~\ref{eq:2} and the luminosity needed to produce 1 neutrino in IceCube. Fig.~\ref{fig:LC} also gives the ``baryon loading'' factor, $\xi$, implied by a given proton luminosity, defined here as $\xi = \vareps_p L_{\vareps_p} / \vareps_{\rm \gamma} L_{\vareps_\gamma}$\footnote{We have approximated $\vareps_{\rm \gamma} L_{\vareps_\gamma} \sim\mathcal{L}_{\gamma}/ \ln{(\rm 320~GeV/100~MeV)}$, where $\mathcal{L}_{\gamma} = 5.66\times 10^{45}~$\erg~
is the $\gamma$-ray luminosity measured with the \fermi-LAT during the 250-day flare.}. 
Considering the long-term average \fermi-LAT flux instead, Eq.~\ref{eq:2} leads to an upper limit on $\vareps_{\nu} L_{\vareps_{\nu}} \approx [6 /(1+Y_{\rm IC})5]\vareps_{\gamma} L_{\vareps_{\gamma}}|_{\vareps^{p\pi}_{\rm syn}} \approx 3 \times 10^{44}~{\rm erg~s^{-1}}$. This is a factor of $\sim 30$ lower than the neutrino luminosity needed to detect 1 neutrino in IceCube, assuming a 10-yr livetime, which is $\mathcal{L}_{\nu}/\ln{(2.6~\mathrm{PeV}/65~\mathrm{TeV})} \approx 8 \times 10^{45}$ \erg. Thus, if the neutrino emission  was related to the long-term emission of \HSPJ, it is easier to satisfy the $\gamma$-ray emission constraint than if the neutrino emission was related to the \fermi-LAT 250 day high-state. These results are also summarised in Fig.~\ref{fig:LC}. Below the threshold for $p\pi$ interactions, protons lose energy via the Bethe-Heitler (BH) process. Unlike in the case of TXS\,0506+056 for \HSPJ there are no observations available to constrain the BH cascade component and the most stringent constraint on the neutrino luminosity comes from the $p\pi$ cascade.
Fig.~\ref{fig:LC}, reveals the difficulty of canonical theoretical models to explain the observation of one neutrino from \HSPJ during the 250 d \fermi\, high state, and to a lesser extent during the 10 yr of IceCube observations. The Poisson probability to detect one neutrino 
is $\sim 0.01$ and $\sim 0.03$ for the two timescales respectively, which could be interpreted as a statistical fluctuation to account for the association. Note, that a similar neutrino luminosity upper limit has been derived in one-zone models of neutrino production of TXS\,0506+056 during its 2017 flare \citep{2018ApJ...863L..10A,2019MNRAS.483L..12C,2019NatAs...3...88G,Keivani:2018rnh,2020ApJ...891..115P}, which must also be interpreted as an upward ($\sim 2\sigma$) fluctuation to account for the observed association. 
On the other hand, Eq.~\ref{eq:2} assumes that neutrinos and $\gamma$-rays are co-spatially emitted. In the presence of multiple emitting zones, and/or an obscuring medium for the $\gamma$-rays, the constraint of Eq.~\ref{eq:2} can be relaxed and larger neutrino luminosity may be produced by \HSPJ (see for example such models for the 2017 flare of TXS\,0506+056: \citealt{2018ApJ...865..124M,2019PhRvD..99f3008L,2019MNRAS.489.4347O,2019ApJ...886...23X,2020ApJ...889..118Z}).

In summary, \HSPJ, with its extremely high \nup\, is the second IBL/HBL which is off the blazar sequence 
 to be detected in the error region of a high-energy neutrino during a flare. 
The Eddington ratio and upper limit to the BLR power we obtained make \HSPJ an unlikely ``masquerading'' BL\,Lac, in contrast to \TXS, pointing to a different class of possible neutrino emitting BL Lac objects, which do not possess a (hidden) powerful BLR but with abundant $>$~keV photons, owing to the high \nup, which may facilitate PeV neutrino production. 
As was the case with \TXS, a possible association points to non-standard (``one-zone'') theoretical models, and/or the existence of an underlying population of sources each expected to produce $\ll 1$ neutrinos in IceCube but with summed expectation $\geq 1$.  Fig.~\ref{fig:LC} reveals that $\sim 150~(30)$ sources identical to \HSPJ are needed to produce one neutrino in 250 days (10 years), corresponding to an expectation of $\lsim$0.01 neutrinos from a single blazar of this type. This would imply that the IceCube sensitivity is still above the expected fluxes from similar individual blazars, and the currently observed neutrino counting, if due to blazars, must be driven by large statistical fluctuations. A possible way to reconcile observations with expectations is to consider that there are about 100 catalogued blazars with properties similar to 3HSPJ095507.1+355101. If each of these objects emits an average flux of $\sim 0.01$ neutrinos in the period considered, we would be in a situation of extremely low counting statistics where the probability of observing one neutrino from a specific blazar is of the order of 1\%. Collectively, however, one neutrino would be expected on similar time-scales from one of the $\sim\,100$ randomly distributed blazars in the underlying population. This scenario is consistent with the current situation where only single neutrino events from each candidate counterparts are observed. Examples supporting this view are the extreme blazars 3HSPJ023248.6+201717, 3HSPJ144656.8-265658, and 3HSPJ094620.2+010452, located inside the 90\% uncertainty region of IC111216A,  IC170506A and IC190819 \citep{Giommi_2020}.

\begin{acknowledgements}
We acknowledge the use of data and software facilities from the ASI-SSDC and the tools developed within the United Nations ``Open Universe'' initiative. 
This work is supported by the Deutsche Forschungsgemeinschaft
through grant SFB\,1258 ``Neutrinos and Dark Matter in Astro and Particle
Physics''. We thank Riccardo Middei for his help with the analysis of NuSTAR data. We thank Matthias Huber and Michael Unger for useful discussions on the interpretation of the IceCube observations. 
\end{acknowledgements}

\bibliographystyle{aa}
\bibliography{3hsp}

\newpage
\section{Appendix}

In this Appendix we give details of the multi-frequency data analysis of 3HSP J095507.1+355101.

\subsection{Swift-XRT}
All Swift-XRT observations were analysed using \swiftdeepsky\, and \swiftxrtproc, the imaging and spectral analysis tools developed within the Open Universe initiative \citep{Giommi2019,Giommi2015}. Both tools are based on the official HEASoft data reduction package, in particular on XIMAGE-4.5 and XSPEC-12, and are particularly useful when analysing a large number of observations, as the tools automatically download the data and calibration files from one of the official archives, generate all the necessary intermediate products, and conduct a detailed standard analysis.
The results of the image analysis are presented in Tab. \ref{imaginggresults} where column 1 gives the observation start time, column 2 gives the effective exposure time, column 3 gives the count rate in the 0.3-10 keV band, and columns 4, 5, 6, and 7 give the flux in the 0.3-10, 0.3-1.0, 1-2 keV, and 2-10 KeV bands, respectively. The largest flux variations are observed in the 2-10 keV band where the intensity varied by over a factor ten, between a minimum of 0.55 $10^{-12}$ \ergs\, on MJD 56233 (Nov. 2, 2012) and a maximum of 6.16 $10^{-12}$ \ergs\, on MJD 58900 (Feb. 21, 2020).

\begin{figure}[h]
\includegraphics[width=0.5\textwidth]{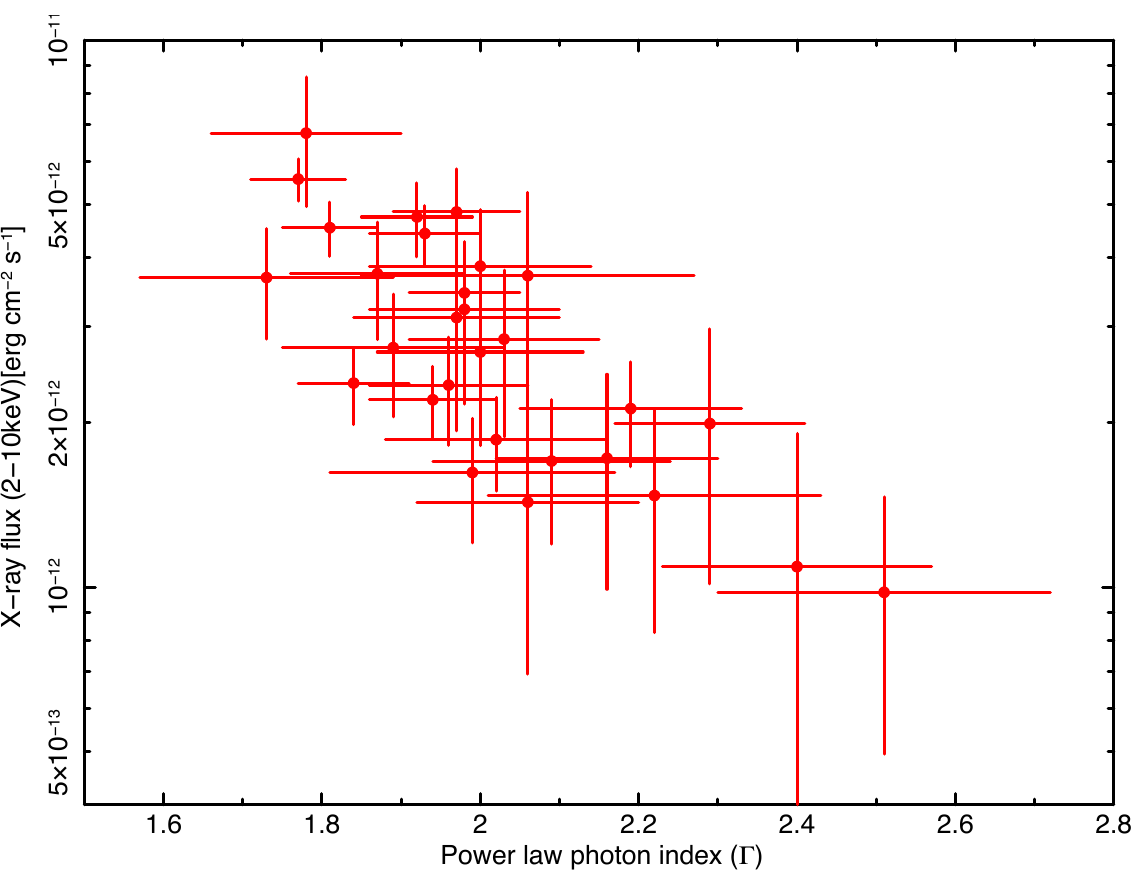}
\caption{The best fit power law spectral index of all the observations with errors smaller than 0.25 is plotted versus the 2-10 keV X-ray flux of 3HSP J095507.1+355101.}
\label{fig:flvsslope}
\end{figure}

Details of the spectral analysis for the cases of power law and log parabola models with NH fixed to the Galactic value, are given in Tab. \ref{spectralresults}. 
Column 1 gives the observation date, 
column 2 gives the best fit photon spectral index with one $\sigma$ error, 
column 3 gives the value of the reduced $\chi^2$ with the number of degrees of freedom (d.o.f.) in parenthesis, columns 4 and 5 give the spectral slope at 1 keV ($\alpha$) and curvature parameter ($\beta$) with one $\sigma$ error, and column 6 
the corresponding reduced $\chi^2$ and d.o.f. Fig.\ref{fig:flvsslope} shows the best fit power law spectral index vs the 2-10 keV flux, for all the  observations where the error on the spectral slope is smaller than 0.25. The figure shows a clear harder-when-brighter trend, a behaviour seen in several other HBL blazars \citep[e.g.][]{Giommi1990}. 
\begin{table*}
\begin{footnotesize}
\begin{center}
\caption{Results of the imaging analysis of all Swift-XRT observations of 3HSPJ095507.1+355101 with exposure time larger than 200 seconds.}
\begin{tabular}{ccccccc}
\hline\hline
Observation & Exposure & Count-rate & Flux  & Flux & Flux & Flux  \\
 start time & time & 0.3-10 keV & 0.3-10 keV & 0.3-1 keV &1-2 keV & 2-10 keV   \\
 MJD & seconds & cts s$^{-1}$& 10$^{-12}$ erg cm$^{-2}$ s$^{-1}$&10$^{-12}$ erg cm$^{-2}$ s$^{-1}$ & 10$^{-12}$ erg cm$^{-2}$ s$^{-1}$& 10$^{-12}$ erg cm$^{-2}$ s$^{-1}$\\
  (1) & (2) & (3) & (4) & (5) & (6) & (7) \\
\hline
  56036.1523 &   845. &  0.13 $\pm$  0.01 &  4.21 $\pm$  0.42 &   1.36 $\pm$  0.20 &   0.83 $\pm$  0.14 &   2.04 $\pm$  0.45 \\
  56208.1641 &  4296. &  0.12 $\pm$  0.01 &  4.09 $\pm$  0.18 &   1.14 $\pm$  0.08 &   0.88 $\pm$  0.06 &   2.04 $\pm$  0.20 \\
  56210.0352 &  3915. &  0.14 $\pm$  0.01 &  4.30 $\pm$  0.19 &   1.44 $\pm$  0.10 &   0.89 $\pm$  0.07 &   1.95 $\pm$  0.20 \\
  56211.8359 &   232. &  0.14 $\pm$  0.03 &  4.81 $\pm$  0.89 &   1.48 $\pm$  0.40 &   1.05 $\pm$  0.29 &   1.02 $\pm$  0.60 \\
  56212.6406 &  2074. &  0.14 $\pm$  0.01 &  4.10 $\pm$  0.25 &   1.50 $\pm$  0.13 &   0.92 $\pm$  0.09 &   1.63 $\pm$  0.24 \\
  56216.0508 &  1559. &  0.10 $\pm$  0.01 &  3.03 $\pm$  0.25 &   1.04 $\pm$  0.13 &   0.72 $\pm$  0.10 &   1.21 $\pm$  0.25 \\
  56217.1172 &   564. &  0.09 $\pm$  0.01 &  3.14 $\pm$  0.46 &   0.89 $\pm$  0.21 &   0.56 $\pm$  0.14 &   1.45 $\pm$  0.48 \\
  56227.0586 &  1447. &  0.12 $\pm$  0.01 &  3.61 $\pm$  0.29 &   1.33 $\pm$  0.15 &   0.76 $\pm$  0.10 &   1.48 $\pm$  0.27 \\
  56230.9336 &   435. &  0.12 $\pm$  0.02 &  4.34 $\pm$  0.64 &   0.94 $\pm$  0.24 &   1.14 $\pm$  0.23 &   1.44 $\pm$  0.57 \\
  56233.0039 &   656. &  0.14 $\pm$  0.01 &  4.81 $\pm$  0.53 &   1.66 $\pm$  0.25 &   1.01 $\pm$  0.17 &   0.55 $\pm$  0.28 \\
  56253.3320 &   809. &  0.14 $\pm$  0.01 &  4.00 $\pm$  0.40 &   2.07 $\pm$  0.27 &   0.60 $\pm$  0.12 &   1.37 $\pm$  0.36 \\
  56254.3320 &  1892. &  0.08 $\pm$  0.01 &  2.39 $\pm$  0.21 &   0.92 $\pm$  0.11 &   0.44 $\pm$  0.07 &   1.02 $\pm$  0.21 \\
  56284.6445 &  1374. &  0.14 $\pm$  0.01 &  3.96 $\pm$  0.31 &   1.73 $\pm$  0.19 &   0.96 $\pm$  0.12 &   1.16 $\pm$  0.26 \\
  56290.2461 &   609. &  0.12 $\pm$  0.01 &  4.45 $\pm$  0.53 &   1.54 $\pm$  0.26 &   0.78 $\pm$  0.16 &   1.20 $\pm$  0.41 \\
  56298.0039 &  1079. &  0.21 $\pm$  0.01 &  6.10 $\pm$  0.41 &   2.26 $\pm$  0.22 &   1.46 $\pm$  0.17 &   2.25 $\pm$  0.39 \\
  56305.0898 &   711. &  0.19 $\pm$  0.02 &  5.73 $\pm$  0.50 &   2.00 $\pm$  0.26 &   1.39 $\pm$  0.18 &   2.22 $\pm$  0.48 \\
  56318.9023 &  1072. &  0.20 $\pm$  0.01 &  6.17 $\pm$  0.43 &   2.38 $\pm$  0.23 &   1.17 $\pm$  0.14 &   2.60 $\pm$  0.42 \\
  56321.9062 &   583. &  0.17 $\pm$  0.02 &  5.93 $\pm$  0.64 &   1.70 $\pm$  0.28 &   0.93 $\pm$  0.18 &   3.39 $\pm$  0.69 \\
  56324.5117 &   346. &  0.18 $\pm$  0.02 &  6.47 $\pm$  0.85 &   1.35 $\pm$  0.31 &   1.38 $\pm$  0.27 &   3.74 $\pm$  0.96 \\
  56329.4531 &  1027. &  0.17 $\pm$  0.01 &  5.47 $\pm$  0.42 &   1.80 $\pm$  0.22 &   1.06 $\pm$  0.14 &   2.61 $\pm$  0.45 \\
  56332.8594 &  1051. &  0.14 $\pm$  0.01 &  4.49 $\pm$  0.37 &   1.54 $\pm$  0.20 &   0.95 $\pm$  0.13 &   1.98 $\pm$  0.38 \\
  56334.9922 &   222. &  0.18 $\pm$  0.03 &  6.34 $\pm$  1.07 &   2.05 $\pm$  0.49 &   1.07 $\pm$  0.31 &   2.36 $\pm$  0.97 \\
  58856.2461 &  2681. &  0.28 $\pm$  0.01 &  9.53 $\pm$  0.37 &   2.70 $\pm$  0.17 &   1.83 $\pm$  0.12 &   5.04 $\pm$  0.41 \\
  58858.7188 &  2523. &  0.24 $\pm$  0.01 &  6.79 $\pm$  0.29 &   2.42 $\pm$  0.15 &   1.88 $\pm$  0.11 &   2.22 $\pm$  0.24 \\
  58859.4336 &  2614. &  0.26 $\pm$  0.01 &  8.39 $\pm$  0.33 &   2.45 $\pm$  0.15 &   1.74 $\pm$  0.11 &   4.16 $\pm$  0.35 \\
  58864.4219 &   373. &  0.30 $\pm$  0.03 &  8.03 $\pm$  0.78 &   3.47 $\pm$  0.49 &   2.31 $\pm$  0.33 &   1.79 $\pm$  0.54 \\
  58869.9922 &   721. &  0.21 $\pm$  0.02 &  6.26 $\pm$  0.54 &   2.12 $\pm$  0.27 &   1.52 $\pm$  0.20 &   2.49 $\pm$  0.51 \\
  58871.7148 &  2328. &  0.29 $\pm$  0.01 &  9.21 $\pm$  0.38 &   2.94 $\pm$  0.19 &   1.92 $\pm$  0.13 &   4.30 $\pm$  0.40 \\
  58873.6406 &  1718. &  0.24 $\pm$  0.01 &  7.57 $\pm$  0.38 &   2.48 $\pm$  0.18 &   1.60 $\pm$  0.13 &   3.44 $\pm$  0.38 \\
  58877.2969 &  1342. &  0.21 $\pm$  0.01 &  6.40 $\pm$  0.39 &   1.99 $\pm$  0.19 &   1.64 $\pm$  0.16 &   2.60 $\pm$  0.39 \\
  58878.0898 &   970. &  0.23 $\pm$  0.02 &  7.76 $\pm$  0.53 &   2.75 $\pm$  0.27 &   1.16 $\pm$  0.15 &   3.97 $\pm$  0.58 \\
  58890.1992 &   724. &  0.25 $\pm$  0.02 &  7.72 $\pm$  0.58 &   2.84 $\pm$  0.31 &   1.58 $\pm$  0.20 &   3.24 $\pm$  0.60 \\
  58895.1562 &   739. &  0.33 $\pm$  0.02 & 11.22 $\pm$  0.76 &   2.81 $\pm$  0.32 &   2.27 $\pm$  0.25 &   6.16 $\pm$  0.83 \\
  58900.0117 &  1978. &  0.28 $\pm$  0.01 &  9.12 $\pm$  0.39 &   2.83 $\pm$  0.19 &   1.88 $\pm$  0.13 &   4.39 $\pm$  0.41 \\
\hline\hline
\end{tabular}
\label{imaginggresults}
\end{center}
\end{footnotesize}

\end{table*}
\begin{table*}[h]
\begin{footnotesize}
\begin{center}
\caption{Results of the spectral analysis of all Swift-XRT observations of 3HSPJ095507.1+355101 with at least 25 net counts.}
\begin{tabular}{ccccccc}
\hline\hline
Observation date & Power law & Reduced $\chi^2$ & Log parabola  & Log parabola & Reduced $\chi^2$ \\
               & $\Gamma$ & & $\alpha$ & $\beta$ & \\
 (1) & (2) & (3) & (4) & (5) & (6) \\
\hline\hline
  56036.1523 &  2.09 $\pm$  0.15 &  1.35 (75) &  2.23 $\pm$  0.19 & -0.48 $\pm$  0.39 &  1.25 (74) \\
  56208.1641 &  1.84 $\pm$  0.07 &  1.15 (227) &  1.75 $\pm$  0.10 &  0.30 $\pm$  0.21 &  1.12 (226) \\
  56210.0352 &  1.94 $\pm$  0.08 &  0.86 (212) &  1.87 $\pm$  0.10 &  0.28 $\pm$  0.24 &  0.83 (211) \\
  56211.8359 &  2.17 $\pm$  0.38 &  0.69 (25) &  1.97 $\pm$  0.41 &  1.58 $\pm$  1.20 &  0.55 (24) \\
  56212.6406 &  1.96 $\pm$  0.10 &  1.10 (155) &  1.92 $\pm$  0.13 &  0.15 $\pm$  0.29 &  1.12 (154) \\
  56216.0508 &  1.99 $\pm$  0.18 &  0.97 (79) &  1.71 $\pm$  0.24 &  1.15 $\pm$  0.57 &  0.83 (78) \\
  56217.1172 &  2.03 $\pm$  0.25 &  0.77 (35) &  1.99 $\pm$  0.32 &  0.15 $\pm$  0.71 &  0.79 (34) \\
  56227.0586 &  2.02 $\pm$  0.14 &  1.11 (103) &  1.89 $\pm$  0.17 &  0.63 $\pm$  0.50 &  1.03 (102) \\
  56230.9336 &  2.17 $\pm$  0.27 &  0.72 (39) &  2.00 $\pm$  0.30 &  0.75 $\pm$  0.81 &  0.73 (38) \\
  56233.0039 &  2.51 $\pm$  0.21 &  1.01 (63) &  2.48 $\pm$  0.22 &  0.52 $\pm$  0.73 &  1.06 (62) \\
  56253.3320 &  2.40 $\pm$  0.17 &  0.94 (77) &  2.44 $\pm$  0.20 & -0.25 $\pm$  0.54 &  0.93 (76) \\
  56254.3320 &  2.06 $\pm$  0.14 &  0.62 (90) &  2.15 $\pm$  0.18 & -0.36 $\pm$  0.42 &  0.59 (89) \\
  56284.6445 &  2.16 $\pm$  0.14 &  0.90 (114) &  1.98 $\pm$  0.16 &  0.92 $\pm$  0.45 &  0.88 (113) \\
  56290.2461 &  2.22 $\pm$  0.21 &  1.12 (58) &  2.08 $\pm$  0.24 &  0.75 $\pm$  0.74 &  1.23 (57) \\
  56298.0039 &  2.29 $\pm$  0.12 &  1.19 (123) &  2.26 $\pm$  0.13 &  0.13 $\pm$  0.39 &  1.19 (122) \\
  56305.0898 &  2.19 $\pm$  0.14 &  0.75 (97) &  2.18 $\pm$  0.18 &  4.47 $\pm$  0.38 &  0.76 (96) \\
  56318.9023 &  2.03 $\pm$  0.12 &  0.94 (132) &  1.92 $\pm$  0.15 &  0.38 $\pm$  0.35 &  0.95 (131) \\
  56321.9062 &  1.73 $\pm$  0.16 &  0.86 (76) &  1.74 $\pm$  0.28 & -2.91 $\pm$  0.53 &  0.87 (75) \\
  56324.5117 &  1.76 $\pm$  0.22 &  0.77 (47) &  1.73 $\pm$  0.28 &  9.21 $\pm$  0.74 &  0.78 (46) \\
  56329.4531 &  2.00 $\pm$  0.13 &  0.84 (113) &  1.94 $\pm$  0.17 &  0.21 $\pm$  0.39 &  0.85 (112) \\
  56332.8594 &  1.89 $\pm$  0.14 &  1.27 (96) &  1.93 $\pm$  0.18 & -0.16 $\pm$  0.44 &  1.28 (95) \\
  56334.9922 &  1.99 $\pm$  0.29 &  0.74 (30) &  2.11 $\pm$  0.31 & -0.83 $\pm$  0.98 &  0.75 (29) \\
  58856.2461 &  1.77 $\pm$  0.06 &  1.04 (274) &  1.70 $\pm$  0.09 &  0.20 $\pm$  0.18 &  1.04 (273) \\
  58858.7188 &  1.98 $\pm$  0.07 &  1.51 (216) &  1.76 $\pm$  0.10 &  0.76 $\pm$  0.21 &  1.37 (215) \\
  58859.4336 &  1.81 $\pm$  0.06 &  0.90 (254) &  1.75 $\pm$  0.09 &  0.18 $\pm$  0.19 &  0.89 (253) \\
  58864.4219 &  2.06 $\pm$  0.21 &  1.00 (70) &  1.80 $\pm$  0.23 &  1.56 $\pm$  0.69 &  0.96 (69) \\
  58869.9922 &  1.97 $\pm$  0.13 &  1.22 (101) &  1.94 $\pm$  0.17 &  8.95 $\pm$  0.38 &  1.23 (100) \\
  58871.7148 &  1.92 $\pm$  0.07 &  1.21 (244) &  1.88 $\pm$  0.09 &  0.15 $\pm$  0.19 &  1.20 (243) \\
  58873.6406 &  1.97 $\pm$  0.08 &  0.93 (188) &  1.97 $\pm$  0.11 & -1.50 $\pm$  0.25 &  0.94 (187) \\
  58877.2969 &  1.87 $\pm$  0.11 &  1.02 (145) &  1.71 $\pm$  0.15 &  0.49 $\pm$  0.31 &  0.95 (144) \\
  58878.0898 &  1.98 $\pm$  0.12 &  1.04 (124) &  1.88 $\pm$  0.16 &  0.34 $\pm$  0.36 &  1.05 (123) \\
  58890.1992 &  2.00 $\pm$  0.14 &  1.21 (113) &  1.80 $\pm$  0.19 &  0.93 $\pm$  0.50 &  1.08 (112) \\
  58895.1562 &  1.78 $\pm$  0.12 &  0.81 (141) &  1.55 $\pm$  0.16 &  0.69 $\pm$  0.36 &  0.76 (140) \\
  58900.0117 &  1.93 $\pm$  0.07 &  1.14 (219) &  1.84 $\pm$  0.10 &  0.28 $\pm$  0.21 &  1.13 (218) \\
 \hline\hline
\end{tabular}
\label{spectralresults}
\end{center}
\end{footnotesize}
\end{table*}

\subsection{NuSTAR}

Data from the NuSTAR observation made shortly after the neutrino arrival were analysed using  the  XSPEC12  package. Photons detected by both telescopes (module A and B), were combined and fitted to spectral models following the standard XSPEC procedure. The source was detected between 3 and 30 keV. A power law spectral model gives a best fit slope of $\Gamma$=2.21 $\pm$ 0.06 with a reduced $\chi^2_{\nu}$ =0.93 with 101 d.o.f. A fit to a log parabola model does not improve the reduced $\chi^2_{\nu}$ and therefore it is not reported here.
A combined fit of the NuSTAR and the quasi-simultaneous Swift-XRT data with a log parabola model gives the following best fit parameters: $\alpha$=1.80$\pm$0.07, $\beta$=0.24$\pm$0.05 for a pivot energy E$_{\rm pivot}$ = 1 keV, and reduced $\chi^2_{\nu}$ = 0.86 with 129 d.o.f.
The corresponding SED peak energy, estimated as  E$_{\rm peak} =10^{(2-\alpha)/2\beta}$ \citep{Massaro2004}, is E$_{\rm peak} \sim$ 2.6 keV.

\subsection{Fermi}
For the analysis of the $\gamma$-ray emission of 3HSP J095507.9+355101 we used the publicly available Fermi-LAT Pass 8 data acquired in the period August 4, 2008 to January 8, 2020 and followed the standard procedure as described in the {\it Fermi} cicerone\footnote{\url{https://fermi.gsfc.nasa.gov/ssc/data/analysis/documentation/Cicerone/}}.
We constructed a model that contains all known 4FGL sources plus the diffuse Galactic and isotropic emissions. In the likelihood fits the normalization and spectral index of all sources within 10$^{\circ}$ (corresponding to the $95$\% \fermi~ point-spread function at 100 MeV) are left free. To calculate an a priori estimate of the required integration time for a significant detection of the source, we used the time-integrated measurement in the \fermi~4FGL catalogue. Assuming a signal dominated counting experiment with  $\chi^2_1$ background test-statistic distribution we know that the median test statistic distribution scales linearly in time $t$, i.e. $\mathcal{TS} \propto t$ and therefore 
\begin{equation}
t_{lc} = (\mathcal{TS}_{lc}\,/\,\mathcal{TS}_{2920d})\cdot 2920 \, [\textrm{days}]
\end{equation}
assuming a quasi-steady emission. Here $\mathcal{TS}_{lc}$ defines the target test-statistic value with required integration time $t_{lc}$. 2920 days and $\mathcal{TS}_{2920d}$ are the live time and significance of the source in the 4FGL catalogue, respectively. Note that in general for the significance $\Sigma = \sqrt{\mathcal{TS}}$. The source is detected with a significance of $5.42 \sigma$ in the 4FGL catalogue and hence the resulting integration times for one and two sigma are 100 and 400 days, respectively. In order to avoid washing out a possible time-dependent signal we chose an integration time of 250 days.

Tab. \ref{sed_points_significance} gives the significance of all \gr\ data points shown in the SED in Fig. \ref{fig:SED}. 
\begin{table*}[]
    \centering
        \caption{Energy dependent significance of the \fermi-LAT spectral energy distribution for the full-mission and the 250 days before the neutrino alert.}
    \begin{tabular}{l|c|c}
        \hline \hline
                            &   &   \\
        Energy Band [GeV] & Full-Mission [$\sigma$]  & MJD 58605.6 - 58855.6 [$\sigma$] \\ 
                            &   &   \\
        \hline \hline 
                    &   &   \\
        0.1 - 0.316 & 0 & 0 \\[1ex]
        0.316 - 1& 1.50  & 0 \\[1ex] 
        1 - 3.16 & 2.82 & 2.15 \\[1ex]
        3.16 - 10 & 4.20 & 1.07 \\[1ex]
        10 - 31.6 & 3.35 & 2.76 \\[1ex]
        31.6 - 100 &  0.& 0 \\[1ex]
        100 - 316 & 2.62 & 0 \\[2ex] \hline\hline
    \end{tabular}
    \label{sed_points_significance}
\end{table*}
\end{document}